\documentstyle[preprint,aps]{revtex}

\begin{document}

\newcommand{\be}{\begin{equation}}
\newcommand{\ee}{\end{equation}}
\newcommand{\bea}{\begin{eqnarray}}
\newcommand{\eea}{\end{eqnarray}}
\newcommand{\ba}{\begin{array}}
\newcommand{\ea}{\end{array}}
\newcommand{\sprime}{^\prime}
\newcommand{\dprime}{^{\prime\prime}}
\newcommand{\tprime}{^{\prime\prime\prime}}

\preprint{\vbox{\hbox{IFP-778-UNC}
\hbox{astro-ph/0002089} 
\hbox{February 2000}
}
}

\title{Relic Neutrinos and Z-Resonance Mechanism
for Highest-Energy Cosmic Rays}
\author{James L. Crooks, James O. Dunn and Paul H. Frampton}

\address{{\it Department of Physics and Astronomy}}

\address{{\it University of North Carolina, Chapel Hill, NC 27599-3255}}

\maketitle

\begin{abstract}
The origin of the highest-energy cosmic rays remains elusive.
The decay of a superheavy particle (X) into an ultra-energetic 
neutrino which scatters from a relic (anti-)neutrino
at the Z-resonance has attractive features.
Given the necessary X mass of $10^{14\sim15}$ GeV, 
the required lifetime, $10^{15\sim16}$ y, renders model-building a 
serious challenge but three logical 
possibilities are considered: 
(i) X is a Higgs scalar in $SU(15)$ belonging 
to high-rank representation, leading to
{\it power}-enhanced lifetime;
(ii) a global X quantum number has 
{\it exponentially}-suppressed symmetry-breaking
by instantons; and (iii) with additional space dimension(s)  
localisation of X within the real-world brane leads to
{\it gaussian} decay suppression, the most 
efficient of the suppression mechanisms considered.
\end{abstract} 

\newpage

The confluence of cosmology and particle phenomenology
benefits both disciplines and can lead to important new
insights.

For protons propagating through the cosmological 
background radiation there
is an energy cut-off, as discussed in {\it e.g.} \cite{FKN},
well-known as the GKZ 
effect\cite{Greisen,KZ}, at
an energy of $E \sim 5 \times 10^{19}$ eV. 
Above this energy, the photoproduction
of pions at the 3-3 resonance 
provides an energy attenuation that prohibits
travel over a distance greater than $\sim 50$ Mpc.

Nevertheless, air showers initiated by a proton (or 
photon) with energies above the GKZ bound have 
been observed\cite{YD}, and this fact needs explanation.

One possibility is that the origin involves the decay of a superheavy particle
as in \cite{FKN} (for related earlier works, see {\it e.g.} \cite{BKV,BS})
but that the decay now produces a high energy neutrino
which scatters from a relic background neutrino at the $Z$ pole
and produces the primary. This Z-burst scenario was suggested in
\cite{Weiler} and further analysed in \cite{GK1,GK2}.

The kinematics of the neutrino-neutrino collision at the
$Z$ pole requires an energy $E_{resonance} = M(Z)^2/(2 M(\nu))$
and taking $M(\nu) = 0.07~eV$ as suggested by the SuperKamiokande
data gives $M_{resonance} \simeq 10^{23}~eV$, just as needed to explain the
data. This is the most attractive feature of the model.

We first estimate the mass and lifetime of the superheavy particle
needed to fit the data. This requires two relationships
derived in \cite{GK2}. Namely the flux of
cosmic rays beyond the GKZ cut-off is estimated as:
\begin{eqnarray}
\Phi_{CR} & = & \frac{C_1}{(4 \pi sr) km^2 (100 y)}  \nonumber  \\
 & \times & \left( \frac{N}{10} \right) 
\left( \frac{\eta}{0.14} \right) 
\left( \frac{\Omega_X}{0.2} \right)
\left( \frac{h}{0.65} \right)^{2}  \nonumber  \\
& \times & \left( B_{\nu} \frac{10^7 t_0}{\tau_X} \right) 
\left( \frac{0.07 eV}{M(\nu)} \right)^{3/2}
\left( \frac{10^{14} GeV}{M(X)} \right)^{5/2}
\label{flux}
\end{eqnarray}
and
\begin{equation}
\frac{\tau_X}{t_0} B_{\nu}^{-1} > C_2  \times 10^5   
\left( \frac{\Omega_X}{0.2} \right) 
\left( \frac{h}{0.65} \right)^2 
\left( \frac{10^{14} GeV}{M(X)} \right)^{3/4}
\label{limit}
\end{equation}
For the numerical dimensionless coefficients 
we find the values $C_1 = 0.33$ and $C_2 = 12.8$
which we use in the following analysis.
The notation is: $N$ is the number of 
protons and photons per annihilation event;
$\eta$ is the relic neutrino density relative 
to the present photon number density
$\eta = (n_{\nu,relic}/n_{\gamma,0})$;
$\Omega_X$ is the contribution of $X$ particles 
to the energy density, relative to the
critical density;
$h$ is the Hubble constant in units of 
$100 km/s/Mpc$; $B_{\nu}$ is the
branching ratio of $X$ into neutrinos;
$t_0$ is the age of the universe; 
$\tau_X$ is the lifetime of $X$; $M(\nu)$ is
the neutrino mass; and $M(X)$ is the mass of $X$.

\bigskip
\bigskip

Assuming central values of all other parameters we 
plot the allowed region of $M(X)$ and $\tau_X$ in Figure 1;
variations in $N, \eta, \Omega_X, h, B_{\nu}, M(\nu)$
can extend the allowed region but here we need
only the order of magnitude estimate.

\bigskip
\bigskip

The value of $M(X)$ must certainly exceed $2E_{resonance}$
so that a two body decay can lead to a neutrino acquiring 
enough energy. Higher energies
can be red-shifted down to $E_{resonance}$ if the 
progenitor $X$ particle
is at a red shift $z > 0$.

The resultant spectrum will cut-off at $M(X)/2$ 
and will be expected to provide a two-component type
of overall spectrum, with a dip around 
$E \sim E_{GKZ}$ as can be seen in the data\cite{YD}.

\bigskip
\bigskip

Since $Z$ decay gives rise to approximately 10 times 
as many photons as nucleons the model predicts
a concomitant number of high-energy photons as cosmic-ray
primaries. Because the data is sparse, it is not 
yet possible to discriminate  
on this basis, as discussed in \cite{Weiler2}; 
this is an important prediction of the Z-burst scenario.

\bigskip
\bigskip

From the above analysis we conclude that the required particle properties
$M(X)$ and $\tau_X$
for the hypothetical state X are well defined in order of magnitude. 
Namely, the
mass M(X) should lie between $10^{14}$ and $10^{15}$ GeV and the lifetime
$\tau(X)$ should lie between $10^{16}$ and $10^{17}$ years.

The remainder of the paper will discuss three possible microscopic
theories or, better, scenarios for this combination of M(X) and $\tau(X)$.
We present these three scenarios in what we regard as their increasing
appeal, from (i) to (iii).

\newpage

(i) {\bf Power suppression.}

The expectation for a particle of this mass is that, unless it is absolutely
stable due to some exact conservation law, it will decay exceedingly quickly
with a lifetime expected to be $\tau \leq 10^{-24}$ seconds.
Since the required lifetime is larger by some 46 or so orders of magnitude,
the longevity is the principal difficulty, as emphasized in \cite{FKN}.

One viewpoint is that this extraordinary suppression of the decay is
an argument against the model, as is the problem, already mentioned, of super-high-energy
photons concomitant with the protons.

Let us here take the viewpoint, as discussed in \cite{Weiler2}
that the photons are a {\it prediction} of the model, to be tested
in future experiments, rather than a fatal flaw. The data on
HECR is probably too sparse to reach any stronger conclusion.

Therefore the only remaining question is longevity.

The first scenario (i) is that considered (in a different model)
in \cite{FKN}. We assume the particle X is a boson and posit a coupling
\[
\frac{g}{M^p} X^{\alpha_1 \alpha_2 .....\alpha_n}_{\beta_1 \beta_2 .....\beta_n}
(\bar{\psi}^{\beta_1}\psi_{\alpha_1})
(\bar{\psi}^{\beta_2}\psi_{\alpha_2})
.........
(\bar{\psi}^{\beta_n}\psi_{\alpha_n})
\]
where the power is $p = 3(n-1)/2$. Let us assume that such a coupling is
gravity-induced and that M is the reduced Planck mass $M \sim M(Pl) \simeq 10^{18} GeV$. 
Then one expects the
lifetime $\tau(X)$ to be of the order of magnitude
\[
\tau(X) \sim (10^{-24} sec.) \times \left( \frac{M(X)}{M(Pl)} \right)^{2p}
\]
in which the mass ratio is $\frac{M(X)}{M(Pl)} \simeq 10^{-3}-10^{-4}$. To arrive at a suppression
of $10^{-46}$ thus requires $2p \sim 12-15$ and $n\sim 5-6$. Thus the X field must
have a high tensorial rank. If this is too much for the reader, skip to
scenario (ii). 

In \cite{FKN} the case $n = 2$ was considered. 
In the spontaneous breaking of $SU(15)$
theory \cite{FLee} such a tensor appears "naturally" 
in the Higgs sector. There is no apparent
need for such a high rank as n = 5 or 6, 
but equally no reason for their absence.
The dimensions of such scalar representations in 
SU(15) are astronomical -
even for n=2 the dimension\cite{FKN,FK} 
is 14,175 while for n=5 and 6 this becomes 
respectively 125,846,784 and 1,367,127,216.   

It is difficult to believe that such power 
suppression could be responsible for
the longevity. It is a logical possibility 
which appears highly contrived.
Thus exponential or gaussian suppression is more appealing.

\bigskip
\bigskip

(ii) {\bf Exponential suppression.}

Let us assume that the superheavy particle X carries a conserved
quantum number $Q_X$ (analagous to baryon number, 
B) and that in perturbation theory
the quantum number is exactly conserved. If there is no open channel
which conserves $Q_X$ then the state will be absolutely stable.

In the case of B in the standard model, it was first shown in 1976 by
't Hooft\cite{Hooft1,Hooft2} that nonperturbative instanton effects
violate conservation and lead to decay of otherwise stable states
such as the proton. The resutant rate is typically exponentially suppressed
by an exponential of the form $exp(-constant/g^2)$ where $g$ is the
gauge coupling constant. Many other examples of such  
suppression are covered in \cite{Shifman}.

Thus, one scenario is that $Q_X$ generates a symmetry of the lagrangian
but $X$ decays with exponential suppression due to instanton effects.
We mention this only for completeness - any quantitative estimation
would require many hypotheses.

\bigskip
\bigskip

(iii) {\bf Gaussian suppression.}

This scenario which is, in our opinion, the most
appealing involves the assumption of at least one extra spatial dimension.
We will take five space-time dimensions, four space and one time.

Let the coordinates be $(x_0, {\bf x}, y)$ with $y$ as the
hypothetical extra dimension on which we now focus.

It used to be thought, up to a decade ago, that any such $y$ must be
compactified at or beyond the GUT scale of $(10^{16} GeV)^{-1}~\sim
~10^{-32}~m$ (recall $1 (GeV)^{-1}~\sim~2 \times 10^{16}~m$).
In 1990, Antoniadis\cite{antoniadis} was the first to
entertain very much larger compactification 
scales $\sim~(1~TeV)^{-1}\sim~10^{-19}~m$. In 1998 it
was pointed out 
\cite{ADD1,ADD2,ADD3,ADD4} that, although the strong
and electroweak interactions of the quarks and leptons need be confined
to a region of $y$ not exceeding $10^{-19}~m$ (the real-world 
brane on which we live), the gravitational interaction
could be decompactified even out to $1~mm~=10^{-3}~m$ without
contradicting experimental data, offering
the possibility of detecting such additional dimensions by deviation
from Newton's Law of Gravity at millimeter scales.

Models in which the fifth dimension contains 
a real-world brane and a suitably separated 
Planck brane which delimits gravitational
propagation have 
been discussed in \cite{verlinde,RS}.
Such models are of interest mainly because they
suggest how to incorporate gravity
in the conformality approach\cite{conf1,conf2,conf3,conf4,conf5,conf6}
which {\it ab initio} describes a flat (gravitationless) space-time.

Let us assume, therefore, that the standard model states are
all confined within a real-world brane with a thickness
of order $10^{-19}~m$ in the $y$ direction. Following \cite{AS}
(see also \cite{MS} and, for the many fold universe, \cite{ADDK})
a scalar field $\Phi(x_{\mu},y)$ which has a $\Phi$
domain wall in the $y$ dimension. In the vicinity of
the wall centered by convention at $y = 0$, 
and with the normalizations of \cite{AS}
the field has the value
\begin{equation}
\Phi(y) = 2 \mu^2 y
\end{equation}

First consider chiral fermions 
(after all, $X$ could be a fermion but we will consider
the boson possibility later). 
In this case we write the five-dimensional action
\begin{equation}
S = \int d^4x dy \bar{\Psi}_i 
[i\gamma_{\mu}\partial/\partial y_{\mu} 
+ \Phi(y) - m_i] \Psi_i + ...
\label{fermion}
\end{equation}
The fermion $\Psi_i$ is now localised at $y = m_i/(2 \mu^2)$.
If the Higgs $H(y)$ is unlocalised inside the domain wall then the 
resulting coupling of $\Psi_i$ to $\Psi_j$ and $H$ has a gaussian
suppression
\begin{equation}
exp ( - C(m_i- m_j)^2)/\mu^2) 
\end{equation}
where C is a coefficient of order unity. 
This is the gaussian overlap of the gaussian tails
of the two wave functions. 
The thickness of the real-world brane is $\tau \sim (\mu)^{-1}$
while the separation of the two fermion wave functions is 
$\sigma \sim (\Delta y)_{ij}$. 
To obtain the required suppression of 
$\sim 10^{-46}$ we need $\tau/\sigma \sim 10$. For example, if
$\tau \sim 10^{-19}~m$, one needs $\sigma \sim 10^{-20}~m$.
Clearly only the ratio $\tau/\sigma$ matters, 
but need not be a large number, in
order to obtain the necessary suppression.

It is worth remarking that this gaussian 
suppression of Yukawa couplings by
localization in the fifth dimension
has a long historical counterpart in the 
localization of states on orbifold singularities
starting with \cite{DFMS,HV,cvetic} in 1987 and subsequent derivative
literature\cite{orb1,orb2,orb3,orb4,orb5,orb6}.
 
The superheavy particle $X$ may not be a fermion 
but a boson, {\it e.g.} the Higgs scalar considered in
scenario (i) above. Fortunately
we can easily extend the localization argument of \cite{AS} to
the case of a boson and obtain a similar result for
the gaussian suppression of the decay amplitude and consequent longevity.
We replace Eq.(\ref{fermion}) by the following:
\begin{equation}
S = \int d^4x dy \phi^{\dagger}_i [\Box - m_S^2 + \Phi(y)^2]\phi_i + .... 
\label{boson}
\end{equation}
and, analagously to the fermion $\Psi$, the boson $\phi$ is localized
around $y = m_S/2\mu^2$ when $\Phi$ again 
has a domain wall centered at $y = 0$.
Identifying $\phi$ with $X$ then provides 
the required longevity by
the same mechanism.
It would be amusing if the highest-energy cosmic
rays provided the first evidence
for an extra spatial dimension!

\bigskip
\bigskip

To summarize the Z-burst mechanism for the highest energy
cosmic rays, it has two positive features:

1) The resonant energy derived from the 
$Z$ mass and the SuperKamiokande neutrino mass is numerically
close to the required energy.

2) The spectrum is predicted to have the two-component shape
suggested by the present data.

On the other hand, there are also two {\it apparently} negative features:

3) The concomitant high-energy photons are not confirmed by present
data. Better data will confirm or refute this important prediction.

4) The longevity of the superheavy particle is such a 
challenge to microscpic model- building
that it may render the model less credible.

\bigskip
\bigskip

It is point (4) which we have attempted to ameliorate in the present article.

\bigskip
\bigskip
\bigskip
\bigskip
\bigskip

{\bf Acknowledgments  } \hspace{0.5cm}

This work was supported in part by the US Department of Energy
under the Grant No. DE-FG02-97ER-41036. We thank 
Ignatios Antoniadis, 
Don Ellison, Markus Luty and Tom Weiler
for discussions.

\bigskip
\bigskip
\bigskip
\bigskip
\bigskip

{\bf Figure 1.}

\bigskip

Allowed region of $M(X)~-~\tau_X$ from Eq.(\ref{flux}) and
Eq.(\ref{limit}) of the text.
In the Figure $a_X = \tau_X(10^7 t_0)^{-1}$ where $t_0$ 
is the age of the universe
and $b_X = M(X)/(10^{14}GeV)$. Variations in 
$N, \eta, \Omega_X, h, B_{\nu}, M(\nu)$ can extend the allowed region
but we use only such order of magnitude estimates.

\end{document}